# Towards an Enabling Environment for Social Accountability in Bangladesh

## Hossain Ahmed Taufiq[1]

### Abstract:

Social accountability refers to promoting good governance by making ruling elites more responsive. In Bangladesh, where bureaucracy and legislature operate with little effective accountability or checks and balances, traditional horizontal or vertical accountability proved to be very blunt and weak. In the presence of such faulty mechanisms, ordinary citizen's access to information is frequently denied, and their voices are kept mute. It impasses the formation of an enabling environment, where activists and civil society institutions representing the ordinary people's interest are actively discouraged. They become vulnerable to retribution. Social accountability, on the other hand, provides an enabling environment for activists and civil society institutions to operate freely. Thus, leaders and administration become more accountable to people. An enabling environment means providing legal protection, enhancing the availability of information and increasing citizen voice, strengthening institutional and public service capacities and directing incentives that foster accountability. Donors allocate significant shares of resources to encouraging civil society to partner with elites rather than holding them accountable. This paper advocate for a stronger legal environment to protect critical civil society and whistle-blowers, and for independent grant-makers tasked with building strong, self-regulating social accountability institutions.

[1] Graduate Teaching Assistant, Public Policy Analysis, Oregon State University, USA & Lecturer (On leave), Global Studies & Governance Program, Independent University Bangladesh. Email: taufiqh@oregonstate.edu , taufiq@iub.edu.bd.

# Introduction

According to Carmen Malena and Mary McNeil, "In the social context, accountability is often defined as the obligation of public power holders to account for or take responsibility for their actions... must explain and justify their actions or face sanctions".[1] There is a general perception that accountability has something to do with answerability. Denis Arroyo points out that accountability is a mechanism that compels powerholders such as public officials to answer for their policies, actions, and use of funds.[2] Sam Hickey and Giles Mohan claim that accountability ensures "that poor or immoral performance is punished in some way."[3]

Existing literature typically refers to two types of accountability mechanisms—horizontal and vertical.[4] Enrique Peruzzotti opines horizontal mechanism guarantees that political, fiscal, administrative and legal checks and balances are present within a state.[5] Generally, horizontal accountability mechanism adopts 'top-down' measures such as anti-corruption ombudsmanship, audits and accounts, legislative accounts committees, human rights watchdogs, and the rule of law. On the contrary, the vertical mechanism is a 'bottom-up' approach, which engages non-state actors such as civil society, citizenship forums, media talk shows, Non-Governmental Organizations (NGOs) and their different social tactics in making the powerholders explain and give reasons for their actions.

Now, the question arises, 'where does social accountability fit within the said mechanisms?' Much has been talked about legal accountability, political accountability, and financial accountability, administrative accountability which are in fact part of either horizontal or vertical mechanisms. Both mechanisms are not considered as magic bullets and have substantial limitations. For instance, in Bangladesh, where bureaucracy and legislature operate with little effective accountability or checks and balances, traditional horizontal or vertical accountability proved to be very blunt and weak.[6] In the presence of such faulty mechanisms, ordinary citizen's access to information is frequently denied, and their voices are kept mute.[7] It impasses the formation of an enabling environment, where activists and civil society organizations (CSO) representing the ordinary people's interest are actively discouraged. They become vulnerable to retribution.

Nevertheless, studies carried out by Tofail Ahmed, Md, Harun or Rashid, Kazi Niaz Ahmmed, and Farhana Razzaque; and Lliya Sumana indicates that social accountability can play a powerful role in filling the gaps left by the horizontal and vertical mechanisms in Bangladesh.[8] Yet, the question remains how and in what way social accountability can fill that gaps. Firstly, this paper seeks to examine the philosophical definition of social accountability and the tools and instruments used within this mechanism. Secondly, the paper presents a primary as well as secondary data interpretation on social accountability to perceive the philosophical necessity and aspects of it. Since 2008, Government, CSO and NGOs in Bangladesh attempted to introduce some paraphernalia indentical to social accountability tools. 'Right to Information Act 2009' and 'Public Procurement Reform -II' of 2008 are among those efforts. This study examines the two initiatives, as well, it inspects Bangladeshi CSO and NGO's capacity in promoting social accountability.

Lack of a theoretical definition: how can accountability social?

Social accountability movement has developed deeply in recent decades. It is a broad term and encompasses key concepts such as democracy, democratic governance, participation and inclusivity. However, it is contested term, and no one has thus far provided a single philosophical definition of social accountability. Thus, different accounts of accountability definition could be found in scholarly pieces connected directly or indirectly to social accountability. Scholars such as Richard Mulgan, Robert D. Behn, and Melvin J. Dubnic categorize it as a many loosely defined political covet such as transparency, responsiveness, responsibility, equity, efficiency, democracy, and governance.[9] Social accountability is an approach, invoked by the civil society or state, refers to promoting good governance by making ruling elites more responsive.[10] John M. Ackerman provides the best conceptual discretion of social accountability in the public sector.[11] According to Ackerman, it is the single most powerful tool to combat three identified threats of good governance, namely corruption, clientelism and capture, and thus can guarantee increased government accountability.[12] It provides an enabling environment for activists and civil society institutions to operate freely. Thus, leaders and administration become more accountable to people. An enabling environment means providing legal protection, enhancing the availability of information and increasing citizen voice, strengthening institutional and public service capacities and directing incentives that foster accountability.

## An urgency to Social Accountability

Without social accountability, the government may fail to provide quality and efficient services such as healthcare, education, water and sanitation, infrastructure, and power services. Social accountability pushes for good governance which is key for development effectiveness, and a decent place among the international community. Social accountability allows a government to understand the needs and demands of its' citizens. In Brazil, social accountability mechanism promoted electoral accountability, enhanced basic public service coverage, and reduced infant mortality; in Uganda, improved health and education outcomes; in India, it ensured less wage theft, and helped people to access ration cards without bribes; in Kenya, it bolstered teaching efforts and improved pedagogy.[13] As stated earlier, in the developing countries, both traditional horizontal (executive, legislature, and judiciary) and vertical mechanism (election) are proved to be weak and blunt in holding public officials accountable. In contrast, social accountability facilitates ordinary citizen's access to information. Through this, they can voice their demands, right to information, and thus ensure accountability of the leaders between elections.

Besides, national and international collaborations in Zambia have shown how effective social accountability can ensure strong citizenship participation and linkages with bureaucrats and politicians.[214] Thus, social accountability tools enhance the possibility of accelerated inclusive development and eradicate poverty through improved public service delivery. In countries like Bangladesh, where the prevalence of misallocation of resources, leakages and corruption are high, social accountability can prove to be beneficial. Citizens have the right to participate in the planning and development of their community and demand better services from their government. A right to information which means government provision of public information can equip the citizens with the knowledge to influence policy and their responsibility. By promoting citizen voices and promoting citizen action, social accountability initiatives can also ensure the empowerment of the marginalized people. It can provide them the necessary information on rights and entitlements, thus can make the public decision-making more transparent and participatory.

Tools and techniques of social accountability:

| Tools | Dimensions |
|---|---|
| **Participatory Budgeting** | Participatory budgeting is a process of directly engaging citizens at different stages and at various capacities in budget formulation and implementation. This is considered as one of the most effective mechanisms of budgeting which operates based on target allocation. With the support of civic engagement and social learning, this mechanism ensures transparency in the actions of public officials and service providers. |
| **The Public Expenditure** | The Public Expenditure Tracking Survey (PETS) is a survey of the public service providers, using a structured questionnaire. In this survey, citizens are |

[2]

| Tools | Dimensions |
|---|---|
| **Tracking Survey (PETS)** | asked about the characteristics of the public facilities, financial flows, service outputs, and accountability related to these services. Alongside with quantitative survey, a qualitative perception study can also be carried out as a complimentary. This tool has proven to be extremely influential in several countries such as Zambia, Uganda, and Peru in order to highlight the use and abuse of public money.[15] |
| **Citizen Report Cards (CRC)** | CRC is a user end feedback system where users can comment and report on the services provided by the public agencies. The entire process is based on a participatory survey, and through the support media and civil society, it can be used as a powerful weapon to ensure public accountability. This mechanism is sought where demand-side data, such as user perceptions on quality and satisfaction with public services are unavailable.[16] The World Bank has recently vouched for the use of report cards in Uganda, Albania, the Philippines and Peru. |
| **Social Audit** | Social Audit tracks and lists the resources available to a service provider. It provides analysis and shares the information to the public through a participatory process. The social audit measures resource use against social purposes. Scope of social audits are varied, and various techniques can be adopted to investigate the government departments. Most common features of SA activities are like evidence-based information producing, awareness raising among the service providers and receivers, improving citizen's access to information. |
| **Citizen Charter** | A Citizen Charter is a documentation that sensitizes citizens about service-related information such as timeline, costs, standards, procedures, and other related issues. This is a need-based design mechanism and can be tailored and listed for different levels of agencies and organizations. It ensures the quality of services by publishing standards. It scales the margin between citizen's expectation from the government service and what they receive. Citizen's charter informs citizens about their rights and entitlements so that they can exercise considerable pressure on service providers to improve their performance. |

| Tools | Dimensions |
|---|---|
| **Public Hearing** | Public Hearing refers to a formal meeting at the community level. The hearing takes place at the presence of local government officials and citizens. In the meeting, participants have the opportunity to exchange information and community affairs. A very good typical output of public hearing is community budgeting. The mechanism empowers citizens to raise their concerns in front of bureaucrats and elected officials, providing meaningful feedback to them regarding citizen's experience and views. |

# Methodology

The study was conducted during November 2018–January 2019. A structured questionnaire was used to reach out to the respondents. Review of published documents on accountability formed the basis of the questionnaire. Articles from reputed journal archives, books, book chapters, conference papers were searched and screened. Both non-governmental and government reports were searched to examine the best practices of Social Accountability in Africa, Latin America, Europe, and Asia.

# Findings

To comprehend the germane issues concerning social accountability, it is essential to examine the discernments among common people regarding access to basic amenities such as finance, education, health, housing and employment. A majority of the respondents (80%) agreed that new job opportunities for advancement in career is extremely important and they placed it in the top priority spot. This has been followed by access to education (78%) and adequate health services (70%).

When asked, 'If a person notices a problem (for example poor drainage, erratic water supply, non-collection of garbage) what will the person will possibly do?', 55% considered calling someone on the phone that 'one had never met before and got their help with the problem'. This means they would like to contact anyone they consider directly or indirectly related to the project in local govt. offices, directorates, bureaus or in ministries. Among the respondents, 80% considered identifying individuals and/or groups to help with the problem, but 85% considered community organization to address the problem. Some also considered calling an elected official or writing an opinion letter to a newspaper.

A trigger for official accountability mechanism:

During the survey, 70% respondents agreed (20% strongly agreed) that social accountability has the potential to trigger the official accountability mechanism. Llyia Sumana's research also corroborates these findings. According to the research, service recipients commonly agreed on the gain of introducing social accountability mechanism and most of them easily uttered common tools such as social audits, performance monitoring, procurement tracking, lifestyle monitoring and citizen report cards, even though they are not from a conscious or advanced class. Perhaps, the pervasive NGO presence in every corner of Bangladesh explains the reasons behind their articulate response and demand for social accountability. Development partner organizations are also actively promoting SA mechanisms in Bangladesh. The World Bank is providing healthy NGO finance to popularize the concept.

However, government service providers hold a very different opinion. They consider the prevailing system of the internal accountability fair, and according to them, it should be preserved. Not only do they reject the SA idea, but also, doubt the competency of ordinary peoples' judgement regarding the performances of the officials. They fear that the officials might be misjudged, and sanctions imposed on them in a prejudiced way. All forms of corruption have two dealing ends—a supply end and a demand end. In Bangladeshi public offices corruption has become systematic and endemic both. For instance, in the land offices, substantial bribing of public officials to secure special allotments of public resources bypassing regulations is a common practice.[17] Non-substantial corruption involves a recurring but modest payment to avoid delays, for controlling (quick or holding) the movement of files, to hold back any important notice, and extinguish any record. Surprisingly corruption triggers efficiency in the public offices, so much so that these are now known as 'speed money'. When it comes to the service provision of a public office, it has become so normal that nobody asks for an alternative.

Enhanced capacity of the civic groups and grassroots people

The foundation of assessing the factors behind citizen's attitudes is the indentation to which a developed civil society exists and how this intermingles with social values and norms to demystify the collective action competences of the question.

Neo-Tocquevillean scholars like Larry Diamond, Robert D. Putnam, Lester M. Salamon, S. Wojciech Sokolowski, and Regina List ponder that civil society is an autonomous, democratic and rich in social capital and civic engagement, tirelessly works for promoting democracy.[3][18]Civil society is responsible for holding public officials accountable. Well-developed civil society is a key piece to complete the SA puzzle. It addresses the regulatory codes, the authority and limitations of public officials. A vigilant civil society can promote democracy and governance in a developing country in a number of ways: "through providing civic education, increasing interest articulation, monitoring the state apparatus and markets, and ensuring better participation and representation of all segments of the society in decision making, aside from the polls."[19] The relationship and gains of civil society and SA mechanism are *vice versa*. Civil societies' ability to carry out advocacy work, their capacity to debate with state and mobilize people, utilize media are essential for successful social accountability movement.[20]

---

[3] De Tocqueville, Alexis. *American institutions and their influence*. AS Barnes, 1873. De Tocqueville (1873) conceived of civil society as a sphere of mediating organizations between the individuals and the state. Neo-Tocquevillean scholars not only argue for the positive link between civil society and democracy but also advocate for building and strengthening civil society in order to build democracy and ensure good governance in third world countries. While de Tocqueville saw civil society as the site of decentralization for democratic governance, neo-Tocquevilleans view civil society as a supporting structure in the state's democratization.

However, in Bangladesh, ordinary citizens are generally less educated, less concerned about their rights and almost entirely unaware of the idea of a modern, democratic state. This leaves them vulnerable to exploitation by the administration, NGOs, and ruling elites. Their frequent absence from acquiring adequate knowledge to avail the citizenry service openings entitled to them has allowed the public service providers to deprive them on service-related issues frequently. Local people are generally grouped, but what they need is a spokesperson standing out for them. CSOs are supposed to take up the role of bringing them in active engagement with the state. However, CSOs failure in doing so is also well-known. This is also quite evident in the survey result. While 23% remained indifferent with the CSO contributions, 50% of respondents disagreed with the notion that civil society can check the power of ruling elites. According to them neither they cannot identify citizen's sufferings adequately, nor CSOs capability in voicing the grievances of the citizens effectively to force the ruling elites or public officials change their corrosive actions.

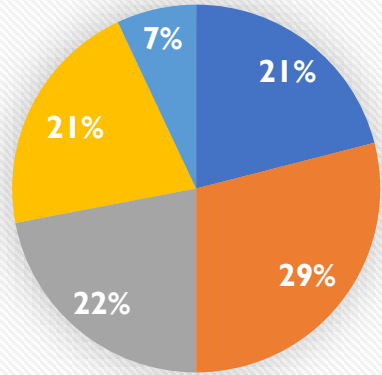

On January 2018, Prime Minister Sheikh Hasina made an appalling comment on the Civil Societies of Bangladesh. She described it as a 'Dustbin of Power and Politics'. She continued, "They're also like that [trash bin] who hang up on their chests 'use me' for politics and power".[21] Such comments on civil societies are not rare globally, but her comments were ironically a harsh truth. Farhat Tasnim's empirical study comprising on case studies upon five CSO examples representing different sectors and levels of the civil society concludes that CSOs in Bangladesh are often politicized and co-opted by different political parties. It is true that NGO contribution in Bangladesh have been remarkable in poverty eradication, literacy and health management through group-based micro credit, rural health awareness systems, and community-based education.[22] In fact, NGO public resource distribution system is different from the nation-state mechanism; in this fashion, Bangladeshi civil society has augmented the state development policy.[23] Yet, very often they also got influenced, polarized, corrupted and ineffective by conflicting political parties.[24] Besides, CSO and NGO developments have reiterated the long patron-client chains involving the very top government leaders to the fringe of Bangladesh; correspondingly, NGO patronage of the poor, allegedly engaged as a new pattern.[25] Despite being ethnically homogenous and casteless, Bangladeshi society is politically polarized and vertically constructed. Unfortunately, instead of spanning social capital and forming ties among different groups, civil society helps exacerbate existing political divisions.[26]

It is high time that Bangladeshi civil societies rethink their strategies. Despite many pitfalls of the Bangladeshi civil societies, their experience, their way of elucidating the socio-economic problems of the citizens, their advocacy techniques and policies would be very much effective and significant to tailor a seamless SA mechanism for a specific locality. In fact, 'it is not essential to always have a skilled and knowledgeable group of people in the demand side of social accountability, rather it is important to participate collectively in an organized way' where civil society can play a lead role in community mobilization.[27]

Right to information and use

Bangladesh introduced the Right to Information (RTI) Act in 2009. According to the RTI preamble, "The act makes free flow of information and people's right to information. The freedom of thought, conscience is recognized in the Constitution...…... shall ensure that transparency, and accountability in all public, autonomous, statuary organizations'.[28] RTI introduction seems mixed. In the survey, when asked having accessed information, citizens have the capacity to use information in actionable ways in Bangladesh, 42% agreed (6% strongly agreed, and 36% agreed), while 36% disagreed (8% strongly disagreed and 29% disagreed).

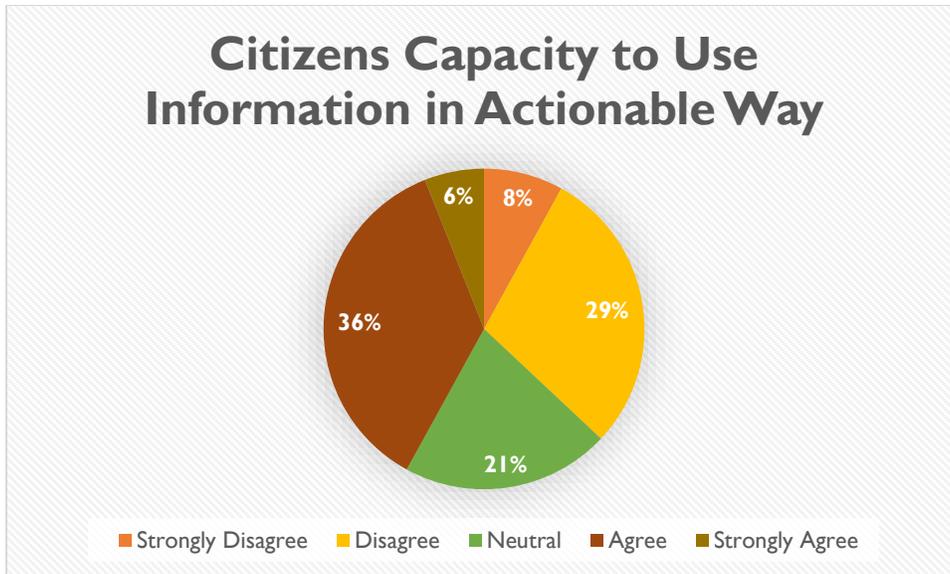

The answer to such variation lies in some of the recent studies. In her study, Sumana argues that Bangladeshi Public Service is and will largely remain unresponsive to the public request for information, and there is a wide possibility that it will trade only the ordinary sort of information and will hold most of the information. Fees charged for information is quite high and unnecessary ancillary costs may have been included. Land offices in Bangladesh keep massive information, and their most valued service is providing authentic information on the record of land titles. For instance, if a citizen, by showing the Official Secrets Act 1923 asks the concerned official, the official may swing around with official rules according to his own sweet will and serve nothing to the inquiring person.[29] If the information were to be passed in accordance with the 2009 RTI Act, many of the perils of people would be resolved. Many land disputes cause from the lost or dented or defective records.

In a different study, Hassan et al., argue that RTI 2009 was very supportive in implementing social accountability practices in the field. 'Private Rural Initiative Program' (PRIP) trust and 'Manusher Jonno Foundation' (MJF) have used Community Score Cards (CSC) in their projects.[30] Though, CSC empowered local citizens to conduct participatory appraisal, forecasting, monitoring and evaluation of basic public community services; on the flip side, the implement1ing agencies capacity lacking, shabby sense of citizenship rights among the local community members, low motivation of service providers and government officials, widespread perception of government resource scarcity among citizens were some of the factors that mired the effective use of CSC as a social accountability tool.[31]

Public Procurement Reform

Every year, Bangladesh spends at-least Tk. 72,000 crores (approximately $8.52 billion) on government procurement.[32] Such substantial investment requires efficient management. If not managed efficiently, it can lead to further expenditures including second-rate output, delays in project implementation and cost overrun. Pondered by this notion, Government launched Public Procurement Reform Project-II (PPRP-II) in 2008 in collaboration with the World Bank.[33] According to the World Bank, the government has invested Tk. 574 crores (approximately $68.10 million) in this project so far and introduced two SA tools as major strategies in this project—i. Electronic procurement and ii. Citizen engagement.[34]

The Central Procurement Technical Unit (CPTU) of Implementation Monitoring and Evaluation (IME) Division, Ministry of Planning, introduced the social accountability initiative in the PPRP II project. The SA features of the project through the assistance of local NGO aim to facilitate citizen engagement in monitoring the execution of public works at the local level. The project deals with the quality assurance of printing of textbook for government primary schools and Local Government Engineering Department's (LGED) public construction works (school and roads). One of the key features of the PPRP social accountability project is that it is being implemented both at the Upazila and Union levels that allow extension of the experiment, which previous was absent in the preceding social accountability initiatives since these were carried out at the UP level. BRAC Institute of Governance and Development (BIGD), an affiliate of the largest NGO BRAC has partnered with CPTU for designing of intervention strategies, advocacy activities and training on different aspects of social accountability and public procurement.

## Conclusion

A traditional accountability model is based on hierarchical and linear relationships. In a democracy, citizens would elect their political representatives who would design and formulate public policies on their behalf. Civil service would implement those policies maintaining appropriate administrative codes. The implementation process will operate within a set of delegations of functions and tiered structure (for example, files will pass from Assistant, Senior Assistant, Deputy, Joint and Additional Secretaries to Secretary level for approval). Political representatives would monitor public officials' performances and expenditure of public money. But this model, as discussed before, has shown its weaknesses and easily susceptible to corruption, embezzlement, and abuse of power. Under such circumstances, civil societies, including NGOs, cannot perform transparently; they act as tools of politics and deviates from their real humanitarian purpose. Social accountability, a new approach has appeared to empower citizen engagement in holding those in power accountable for their actions and decisions. By not undermining or replacing the role of traditional accountability mechanism, social accountability mechanisms are to be used to identify the oversights of the former and develop strategies to improve the performance of the officials, rather than to impugn or punish local officials or even fix their responsibilities. The mechanisms offer a feedback system to the citizens, where if performance is unsatisfactory, officials are pushed to fulfilling their obligations. It engages in citizens problem-solving and helps to create mutual trust between them and public officials. Also, it encourages civic participation in decision-making processes at local levels, raises community awareness, promotes skill or capacity of service beneficiaries and stakeholders; run advocacy for enhanced policy agenda for service provision.

Globally, the need for social accountability arose mainly from as responses to home-grown power imbalances and a longing to improved services. Social, political, and economic contexts have much to do in deciding the risks and benefits of the application of a particular social accountability mechanism. Therefore, social accountability cannot be considered as a silver bullet or quick remedy. Typically, social alertness and movement originate in civil society and in the public spheres such as social media (Facebook, Twitter, Instagram, etc.), and these are foundations of social accountability. But, apart from a few exceptions, these platforms thus far failed to generate the conducive environment for an active social accountability mechanism.

Nevertheless, projects such as PPRP II has shown promising applications of social accountability initiatives. Ultimate success can be achieved in SA through the successful interactions of the citizens, bureaucrats and the state. Road to such intermingling may not be easy; disagreement and doubt will happen but a determined approach can lead to effective state-society interplay. In such a tedious process, the right interface and balance between the leadership of the state and the civil society are mandatory.

The accomplishments of SA tools of India, Cambodia, Philippines, Uganda, Kenya, and Brazil can be a motivation for Bangladesh. However, those tools while applying it in our own context may not produce the same result. Some of the tools may deem less threatening to the public officials and ruling elites than the others will. NGOs and CSOs can be engaged to design and execute initiatives such as Citizen Report Card and apply in broader or central level. On the other hand, service providers and service users can design and run Community Score Cards. While citizenship report cards are applicable in macro settings, Community Score Cards are appropriate for community-level civic engagement and policy designing. Each mechanism can show their strengths and weaknesses depending on the context where they are applied. Public dealings such as land, health services, agricultural credit, education tend to be more favorable for the NGOs and grassroots organizations than the government to act as a facilitator. However, if the government opening up the spaces first and civil society joining later seems more practical. Citizens will be more willing to participate if they discover that they are contributing to the government activities and if dialogue opportunity with honest and sincere public servants are being facilitated. The errands of endorsing and accomplishing social accountability must be done within the context of political actions and in the presence of active civil society. Though social accountability remains unfathomable, lingering, given the confinements of Bangladeshi political culture, bureaucracy, and donor dependency for financial resources yet it is vital for attaining good governance.

---

[1] Carmen Malena, and Mary McNeil, "Social accountability in Africa: An introduction." *Demanding Governance* (2010): 1.

[2] Dennis Arroyo, "Summary Paper on Stocktaking of Social Accountability Initiatives in Asia and Pacific (draft)." *Washington, DC: World Bank* (2004).

[3] Sam Hickey, and Giles Mohan, "The politics of establishing pro-poor accountability: what can poverty reduction strategies achieve?." *Review of international political economy* 15.2 (2008): 234-258. Page: 236.

[4] World Bank, Social Accountability in the Public Sector A Conceptual Discussion and Learning Module,
The International Bank for Reconstruction and Development/The World Bank, USA, 2005, from:
https://siteresources.worldbank.org/PUBLICSECTORANDGOVERNANCE/Resources/AccountabilityGovernance.pdf, last consulted February 12, 2019.

[5] Enrique Peruzzotti, "The workings of accountability: contexts and conditions." *From Inertia to Public Action* (2011): 53.

[6] Tofail Ahmed, Md Harun Or Rashid, Kazi Niaz Ahmmed, and Farhana Razzaque, "Social Accountability Mechanisms: A Study on the Union Parishads." BRAC Institute of Governance and Development, BRAC University, 2016. from:

https://www.researchgate.net/publication/322852588_Social_Accountability_Mechanisms_A_Study_on_the_Union_Parishads_in_Bangladesh, last consulted October 15, 2018.

[7] Mario Claasen, and Carmen Alpín-Lardiés, eds. *Social Accountability in Africa: Practitioners' experiences and lessons*. African Books Collective, 2010.

[8] Tofail Ahmed et al. op. cit; Lliya Sumana, *Social accountability and the quality of service in public offices of Bangladesh*. Diss. BRAC University, 2009.

[9] Richard Mulgan, "'Accountability': an ever-expanding concept?" *Public administration* 78.3 (2000): 555-573; Robert D. Behn, *Rethinking democratic accountability*. Brookings Institution Press, 2001; and Melvin J. Dubnick, "Seeking salvation for accountability." *annual meeting of the American Political Science Association*. Vol. 29. 2002.

[10] Badru Bukenya, Sam Hickey, and Sophie King, "Understanding the role of context in shaping social accountability interventions: towards an evidence-based approach." *Manchester: Institute for Development Policy and Management, University of Manchester* (2012)

[11] Ackerman, John. "Co-governance for accountability: beyond "exit" and "voice"." *World Development* 32.3 (2004): 447-463.

[12] Ibid.

[13] Jonathan A. Fox, "Social accountability: what does the evidence really say?." *World Development* 72 (2015): 346-361.

[14] "Government and citizens come together for social accountability commitment." Counterpart International, October 10, 2017, from: https://www.counterpart.org/stories/saszambia-zambian-government-citizens-social-accountability-commitment/, last consulted August 16, 2019

[15] Tofail Ahmed et al. op. cit.

[16] World Bank op. cit.

[17] Sumana op. cit.

[18] Larry Diamond, *Developing democracy: Toward consolidation*. JHU Press, 1999; Robert D. Putnam, *Bowling alone: The collapse and revival of American community*. Simon and Schuster, 2001; Robert D. Putnam, *Bowling alone: The collapse and revival of American community*. Simon and Schuster, 2001; Lester M. Salamon, S. Wojciech Sokolowski, and Regina List. *Global civil society: An overview*. Baltimore, MD: Center for Civil Society Studies, Institute for Policy Studies, The Johns Hopkins University, 2003.

[19] Farhat Tasnim, "Politicized civil society in Bangladesh: Case study Analyses." *Cosmopolitan Civil Societies: An Interdisciplinary Journal* 9.1 (2017): 99.

[20] Ibid.

[21] "Bangladesh civil society a dustbin of politics and power." *Dhaka Tribune,* August 16, 2019, from: https://www.dhakatribune.com/bangladesh/2018/01/24/bangladesh-civil-society-dustbin-politics-power, last consulted October 17, 2019;

[22] AKM Ahsan Ullah, and Jayant K. Routray. *NGOs and Development, Alleviating Rural Poverty in Bangladesh*. Book Mark International, 2003; Ruhul Amin, *Development Strategies and Socio-Demographic Impact of Non-Governmental Organizations: Evidence from Rural Bangladesh*. University Press Limited, 1997; Asif Dowla, "In credit we trust: Building social capital by Grameen Bank in Bangladesh." *The Journal of Socio-Economics* 35.1 (2006): 102-122.

[23] Ken'ichi Nobusue, "A Large NGO Sector Supported by Foreign Donors." *The State and NGOs: Perspective from Asia* 25 (2002): 34.

[24] Fahimul Quadir, "How "civil" is civil Society? Authoritarian state, partisan civil society, and the struggle for democratic development in Bangladesh." *Canadian Journal of Development Studies/Revue canadienne d'études du développement* 24.3 (2003): 425-438.

[25] David Lewis, "On the difficulty of studying 'civil society': reflections on NGOs, state and democracy in Bangladesh." *Contributions to Indian sociology* 38.3 (2004): 299-322; Sarah C. White, "NGOs, civil society, and the state in Bangladesh: The politics of representing the poor." *Development and change* 30.2 (1999): 307-326; S. Hashemi, 1995. NGO accountability in Bangladesh: Beneficiaries, donors and the state. *NGOs: Performance and accountability: Beyond the magic bullet*, pp.103-110.

[26] Quadir op. cit.

[27] Sumana, op. cit.

[28] "The Right to Information Act, 2009, Bangladesh." Bangladesh Gazette, 6 April, 2009, from: http://www.humanrightsinitiative.org/programs/ai/rti/international/laws_papers/bangladesh/bangladesh_rti_act_2009_summary.pdf, last consulted October 21, 2019.

[29] Sumana op. cit.

[30] Hassan, Mirza M., Sayeda Salina Aziz, and Nadir Shah, "Social accountability in public procurement: how citizen engagement can make a difference." (2016).

[31] ANSA. "Can Community Score Cards Make a difference-The case of Bangladesh, ANSA." 2012.

[32] "How e-GP save taxpayers tens of billions each year." The Daily Star, April 13, 2016, from: http:// www.thedailystar.net/op-ed/politics/how-e-gpsave- taxpayers-tens-billions-each-year-1208140, last consulted September 11, 2018.

[33] The World Bank. World Bank Helps Expand Electronic Public Procurement in Bangladesh, 2016[Online], from https://www.worldbank.org/en/news/press-release/2016/07/25/world-bank-helps-bangladesh-expand-electronic-public-procurement-system , last consulted September 11, 2016.

[^34] Ibid.